\documentclass[prb,twocolumn,a4paper,superscriptaddress,aps,floatfix]{revtex4-1}

\usepackage[utf8]{inputenc}
\usepackage{amsmath,amsfonts,amssymb}
\usepackage{graphicx,graphics}
\usepackage{epstopdf,csquotes,}
\usepackage{color,textcomp,bm}
\usepackage{subfigure,array,xspace} 
\usepackage[squaren,Gray,cdot]{SIunits}
\usepackage{float}
\usepackage[pdfstartview=FitV,bookmarks=true, linktocpage=true,colorlinks=true,
            urlcolor=blue,linkcolor=red, citecolor=blue]{hyperref}   
\usepackage{mathdots}
\usepackage{MnSymbol}
\usepackage{placeins}
\usepackage{multibib}
\usepackage[squaren,Gray]{SIunits}
            
\graphicspath{{Figures/}}
\begin{abstract}
This paper provides a theoretical investigation of negative refraction and focusing of elastic guided waves in a free-standing plate with a step-like thickness change. Under certain conditions, a positive phase velocity (forward) Lamb mode can be converted into a negative phase velocity (backward) mode at such interface, giving rise to negative refraction. A semi-analytical model is developed in order to study the influence of various parameters such as
the material Poisson's coefficient, the step-like thickness, the frequency and the incidence angle. To this end, all the Lamb and shear horizontal propagating modes, but also a large number of their inhomogeneous and evanescent counterpart,s are taken into account. The boundary conditions applied to the stress-displacement fields at the thickness step yields an equation system. Its inversion provides the transmission and reflection coefficients between each mode at the interface. The step-like thickness and Poisson's ratio are shown to be key parameters to optimize the negative refraction process. In terms of material, Duralumin is found to be optimal as it leads to a nearly perfect conversion between forward and backward modes over broad frequency and angular ranges. An excellent focusing ability is thus predicted for a flat lens made of two symmetric thickness steps. Theoretical results are confirmed by a numerical FDTD simulation and experimental measurements made on an optimized Duralumin flat lens by means of laser interferometry. This theoretical study paves the way towards the optimization of elastic devices based on negative refraction, in particular for cloaking or super-focusing purposes.

\end{abstract}

\begin{document}

\title{Negative refraction of Lamb modes: A theoretical study}

\author{François Legrand, Benoît Gérardin, Jérôme Laurent, Claire Prada, Alexandre Aubry}\email{alexandre.aubry@espci.fr}
\affiliation{ESPCI ParisTech, PSL Research University, CNRS, Institut Langevin, UMR 7587, 1 rue Jussieu, F-75005 Paris, France}
\date{\today}

\maketitle

\section{Introduction}

Negative refraction has drawn a considerable attention for the last twenty years whether it be for wave focusing\cite{pendry2000negative}, lensing \cite{veselago1968}, imaging \cite{fang2005sub}, or cloaking \cite{leonhardt2006optical} purposes. In a negative index material, the energy flow as dictated by the Poynting vector is in the opposite direction to the wave vector. \cite{veselago1968} This peculiar property implies that, at an interface between positive and negative index material, waves are bent the unusual way relative to the normal. Any negative refracting slab thus forms a flat lens which does not suffer from any spherical aberration \cite{pendry2000negative}. Negative refraction has also given rise to the notion of complementary media and the ability to cancel the propagation of waves by adjoining two mirror regions of opposite refractive indices \cite{pendrycomp}.

Most experiments on negative refraction of elastic waves have been achieved either using phononic crystals,\cite{morvan2010experimental,croenne2011negative,pierre2010negative,dubois2013flat} or meta-materials,\cite{zhu2014negative} an arrangement of tailored sub-wavelength building blocks from which the material gains unusual macroscopic properties. Nevertheless, these man made materials often rely on resonating structures, a feature that induces strong energy dissipation losses. More recently, an alternative way has been explored for elastic guided waves. An elastic plate actually supports an ensemble of modes, the so-called Lamb waves, which exhibit complex dispersion properties. Interestingly, some Lamb modes, often referred to as backward modes, display a negative phase velocity.\cite{tolstoy1957wave,mindlin1960waves,meitzler1965backward,negishi1987existence}
This particularity comes from the repulsion between two dispersion branches with close cut-off frequencies, corresponding
to a longitudinal and a transverse thickness mode of the same symmetry. The lowest branch exhibits a minimum corresponding
to a zero-group velocity (ZGV) point.\cite{prada2005laser,holland2003air} This peculiar property has been taken advantage
of to achieve negative refraction\cite{bramhavar2011negative,philippe2015focusing} through mode conversion between forward and backward propagating modes (or vice versa) at a step-like thickness discontinuity.

In this paper, we investigate theoretically, the conversion of propagating modes at a thickness step in order to optimize the negative refraction effect. This problem has already been studied for normal \cite{schaal2016lamb,poddar2016scattering} and oblique incidence \cite{feng2016scattering} at frequencies that only imply low-order modes and do not involve any backward mode. Following the  approach of a recent study on negative reflection of Lamb waves at a free plate edge,\cite{gerardin2016negative} we  develop a semi-analytical model to calculate the reflection and transmission coefficients between Lamb modes at a symmetric step discontinuity. The optimal parameters (Poisson's ratio, material, thickness ratio) to reach an efficient negative refraction effect over a broad angular range and a wide frequency bandwidth are then determined using this model. Theoretical results are ultimately confirmed by means of both an FDTD numerical simulation and an ultrasound laser experiment performed on a plate whose design has been priorly optimized.

\section{Determination of the Plates Modes}
Let us consider a homogeneous isotropic plate of thickness $d=2h$. We first derive the plate modes in the right-handed system $(x_1,x_2,x_3)$ whose $x_1$-axis is the propagation direction of the wave and $x_2$-axis is the normal to the plate. The displacement field $\boldsymbol{u}=(u_1,u_2,u_3)^T$ and the stress tensor $\boldsymbol{\sigma}=[\sigma_{ij}]$ obeys the elasticity equations given by :
\begin{equation}\label{eq:elasticityequation}
-\rho \omega^2 \boldsymbol{u}=\nabla \cdot \boldsymbol{\sigma},
\end{equation}
where $\rho$ is the density of the material, and $\omega$ is the pulsation.
The boundary conditions correspond to the cancellation of the stress tensor on the plate surfaces, $\boldsymbol{\sigma} \cdot \boldsymbol{n}=0$, where $\boldsymbol{n}$ is the normal to the surface boundary. Considering the geometry of the problem as shown on Fig \ref{fig:stepschema_curve}.a, solutions are in the form
\begin{equation}
\left\lbrace u_i(x_1,x_2),\sigma_{ij}(x_1,x_2)\right\rbrace=\left\lbrace u_i(x_2),\sigma_{ij}(x_2)\right\rbrace \cdot e^{(i kx_1)}. \nonumber
\end{equation}
Two sets of solutions satisfy these equations : shear horizontal (SH) modes that are polarized orthogonally to the propagation plane ($u_1=u_2=0$) and Lamb modes, polarized in the propagation plane ($u_3=0$). Both families are composed of an infinite number of modes, called propagating, evanescent or inhomogeneous for a real, pure imaginary or complex wave number $k$ respectively.
Both Lamb and SH modes can be separated in two independent families of symmetrical and antisymmetrical modes. Symmetric/antisymmetric SH modes correspond to an even/odd $u_3(x_2)$ polarization along the plate thickness. Symmetric/antisymmetric Lamb modes have an even/odd in-plane component $u_1(x_2)$ combined with an odd/even transverse component $u_2(x_2)$.

We briefly recall the equations for SH and Lamb modes that are fully described in various textbooks\cite{royer1996ondes,achenbach2012wave}.

\subsection{SH modes}

The well known SH mode dispersion is
\begin{equation}
\frac{\omega^2}{c_T^2}-k^2=\left( \frac{n\pi}{2h} \right)^2,
\label{disprelaSH}
\end{equation}
with $c_T$ the shear wave velocity and $n=0,1,2...$.
The corresponding displacement field is:
\begin{equation}
u_3(x_2)=\cos\left(\frac{n\pi}{2h}(x_2+h)\right).
\label{u3SH}\nonumber
\end{equation}
The stress field is then expressed from the displacement field using Eq. \eqref{eq:elasticityequation}.

\subsection{Lamb modes}
Symmetrical Lamb modes are solutions of the following dispersion relation, often referred to as the Rayleigh-Lamb equation\cite{royer1996ondes,auld1973acoustic}
\begin{equation}
\frac{\omega^4}{c_T^4}=4k^2q^2\left\lbrace 1-\frac{p}{q} \frac{\tan(ph)}{\tan(qh)}\right\rbrace,
\label{Rayleighlambequation}
\end{equation}
with $\omega$ the pulsation, $p^2={\omega^2}/{c^2_L}-k^2$, $q^2={\omega^2}/{c^2_T}-k^2$, $c_L$ the longitudinal wave velocity, $c_T$ the shear wave velocity. 

At a fixed pulsation $\omega$, a discrete set of wave numbers $k_n$ satisfy Eq. \eqref{Rayleighlambequation}, and only a finite number of propagating modes are supported by the plate, whereas it exists an infinite number of evanescent or inhomogeneous modes. The components of the displacement field for each mode can be expressed as follows\cite{royer1996ondes} :
\begin{eqnarray}
\begin{array}{l}
u_1^{(n)}(x_2)=-\left[k_n\cos(p_nx_2)-R_nq_n\cos(q_nx_2)\right],\\
u_2^{(n)}(x_2)=i\left[p_n\sin(p_nx_2)+R_nk_n\sin(q_nx_2)\right],
\end{array}\nonumber
\label{fieldexpression}
\end{eqnarray}
with 
\begin{equation}
R_n=\frac{(k_n^2-q_n^2)\cos(p_nh)}{2k_nq_n\cos(q_nh)}.\nonumber
\end{equation}
Again, the stress field can be deduced from $\mathbf{u}$ using Eq.\eqref{eq:elasticityequation}.

\subsection{Dispersion Curves}
\begin{figure*}[tb]
  \centering
   \includegraphics[width=\textwidth]{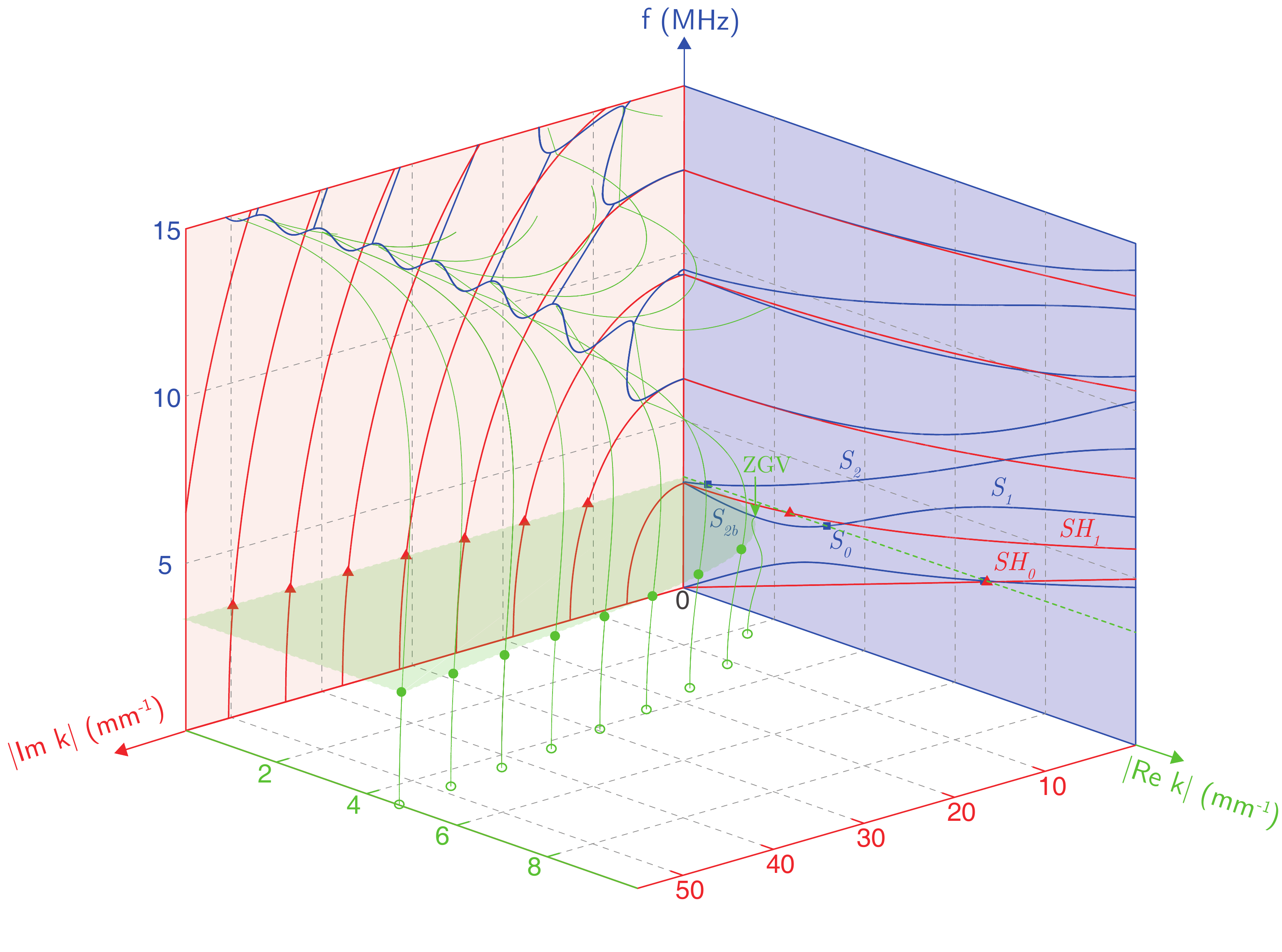}
  \caption{\label{fig:dispcurve3d} Dispersion curves of elastic guided modes in a $1$ mm thick Duraluminium plate computed from Eqs. \eqref{disprelaSH}-\eqref{Rayleighlambequation}: propagating and evanescent SH modes (red), propagating and evanescent Lamb modes (blue) and inhomogeneous Lamb modes (green).}
\end{figure*}

As the thickness step is symmetrical with respect to the $x_2 = 0$ plane, the scattering at a free edge preserves the mode’s symmetry. As a consequence, a symmetrical Lamb mode is reflected and transmitted into symmetrical Lamb and SH modes. In the following we will only consider symmetrical modes. Figure \ref{fig:dispcurve3d} displays the dispersion curves of the SH and Lamb modes deduced from Eqs.[\eqref{disprelaSH},\eqref{Rayleighlambequation}] for a duralumin plate ($\rho=2790$ kg/m$^3$, $c_L=6.4$ mm/$\mu$s, $c_T=3.1$ mm/$\mu$s).
The symmetric zero-order Lamb mode $S_0$ is the extensional mode of the plate. It exhibits free propagation to zero frequency, whereas the higher order modes admit a cut-off frequency. In particular, the $S_1$ and $S_2$ modes have cut-off frequencies at $f=V_T/d$ and $f=V_L/2d$, corresponding to shear and longitudinal thickness resonances, respectively. One peculiar property of Lamb waves is the existence of branches for which phase velocity $\omega/k$ and group velocity $\partial \omega / \partial k$ are of opposite sign. The corresponding modes, often referred to as backward modes, naturally display a negative phase velocity. They originate from the repulsion between two dispersion branches having close cut-off frequencies, corresponding to a longitudinal and a transverse thickness mode of the same symmetry. This is the case for $S_1$ and $S_2$ modes displayed in Fig. \ref{fig:dispcurve3d} in the case of a Duralumin plate. The lowest branch $(S_1)$ exhibits a minimum corresponding to a zero-group velocity (ZGV) point\cite{prada2005laser,holland2003air}. Above this resonance, there is a coexistence of a negative phase velocity (backward)  $S_{2b}$-mode and a positive phase velocity (forward) $S_1$-mode. 

\section{Problem's geometry and equation system}

\begin{figure}[tb]
  \centering
  \includegraphics[width=\columnwidth]{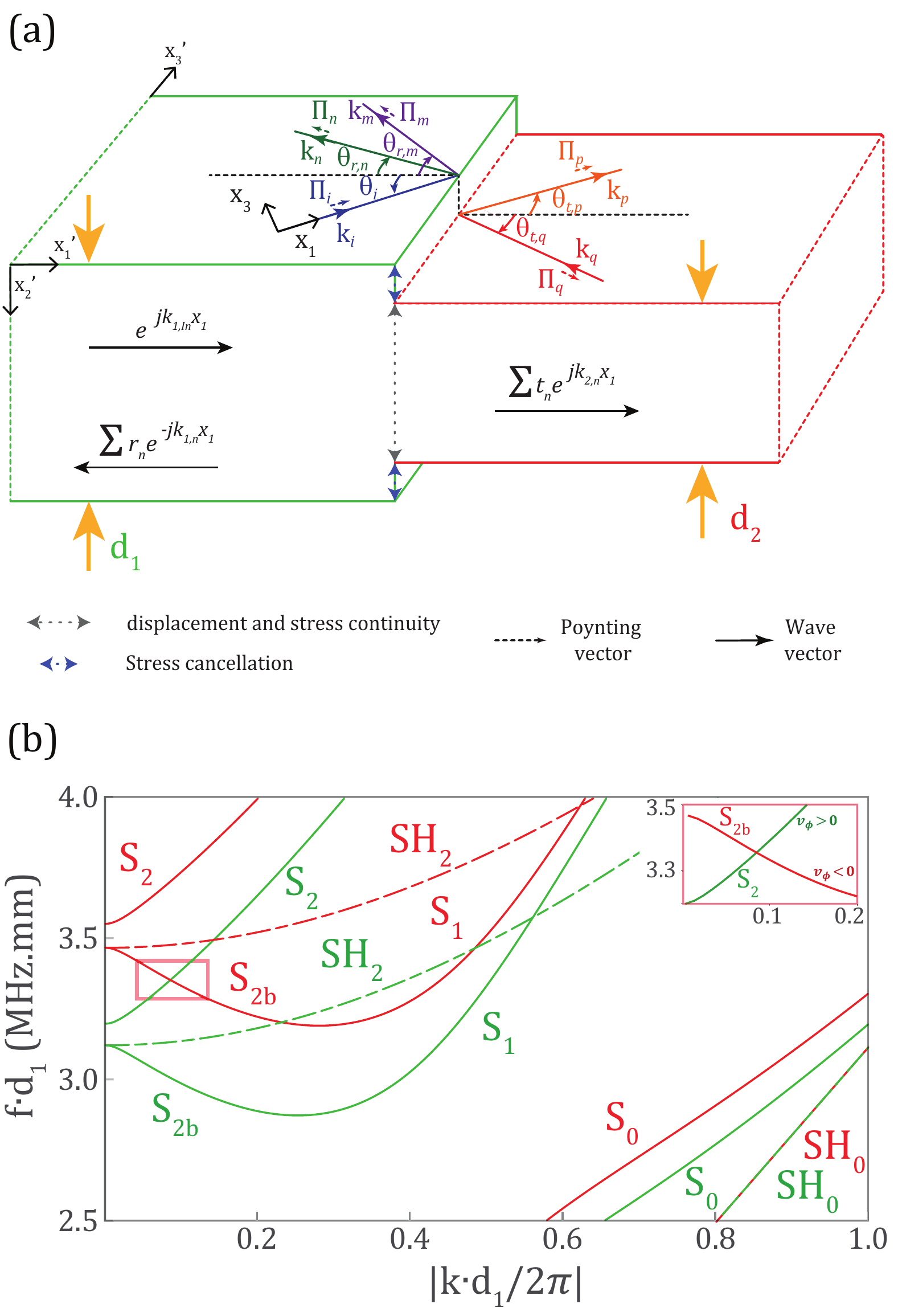}
  \caption{\label{fig:stepschema_curve} (a) Geometry of the problem that shows the interaction of a Lamb mode with a step interface. To satisfy the continuity equations at the interface the incident wave is transmitted and reflected into an infinite combination of Lamb and SH modes.(b) Lamb and SH propagating modes dispersion curves in both parts of the plate (green for $d_1=1$ mm and red for $d_2=0.9$ mm). The forward propagating mode $S_2$  in the thick part intersects the backward propagating mode $S_{2b}$ in the thin part.}
\end{figure}

As shown in previous studies\cite{bramhavar2011negative,philippe2015focusing}, negative refraction of Lamb waves can be achieved by conversion of a forward mode into a backward mode at a thickness step. For the sake of simplicity, we will here consider the conversion between the forward and backward modes $S_2$ and $S_{2b}$ at a symmetric step (Fig. \ref{fig:stepschema_curve}a). Such a geometry is actually optimal for symmetric modes. 

To study the interaction of a mode of oblique incidence, a second right-handed system $(x'_1,x_2,x'_3)$ is introduced [see Fig. \ref{fig:stepschema_curve}(a)]. The axis $x'_1$ is oriented along the step interface while the axis $x'_3$ is normal to this step. The thickness is supposed to be $d_1 = 2h_1$ for $x'_1<0$ and $d_2 = 2 h_2$ for $ x'_1>0$ with $h_1>h_2$. We then consider an incident wave coming from the thick part.

Figure \ref{fig:stepschema_curve}(b) displays the dispersion curves of the Lamb and SH modes in each part of the plate. Right-going (respectively, left-going) propagating modes correspond to a positive (respectively, negative) group velocity $\partial \omega /\partial k$, whereas the evanescent and inhomogeneous right-going (respectively,left-going) modes correspond to wave numbers with strictly positive (respectively, negative) imaginary parts. Because the dispersion curves scale with the plate thickness [Eqs. \eqref{disprelaSH} and \eqref{Rayleighlambequation}], the forward propagating mode $S_2$ in the thick part ($x'_1<0$) crosses the backward propagating mode $S_{2b}$ in the thin part of the plate. As already observed experimentally\cite{philippe2015focusing,bramhavar2011negative}, this crossing point gives rise to negative refraction though an efficient conversion between these two modes at a thickness step. This conversion is now investigated in details. 

Let us consider an incident right-going $S_2$ mode, of wave number $k_I$, carrying an unit energy flux towards the step with an angle of incidence $\theta_I$,  with respect to the axis ($x'_1$), as depicted in Fig. \ref{fig:stepschema_curve}(a). The corresponding stress displacement field is denoted as $\left\lbrace {u'}_i^{I},{\sigma'}_{i,j}^{I}\right\rbrace$. In order to satisfy the stress-free condition at the interface, this incident Lamb mode is reflected/transmitted into an infinite combination of left-going/right-going Lamb and SH modes of wave numbers $k_{r,n}$ and $k_{t,n}$, respectively. It is necessary to consider not only the propagating modes but also the different evanescent and inhomogeneous Lamb and SH modes. For a given mode $n$, the reflection and transmission angles, $\theta_{r,n}$ and $\theta_{t,n}$, are determined by the conservation of the component of the wave vector along $x'_3$, $k_I \sin (\theta_I)=k_{r,n} \sin (\theta_{r,n})= k_{t,n} \sin (\theta_{r,n})$.
The displacement-stress fields,$\left\lbrace \tilde{u}_i^{'(n)},\tilde{\sigma}_{i,j}^{'(n)}\right\rbrace$ and $\left\lbrace {u}_i^{'(n)},{\sigma}_{i,j}^{'(n)}\right\rbrace$, of respectively reflected and transmitted modes expressed in the coordinate system $(x'_1,x_2,x'_3)$, can be obtained from the displacement-stress
fields $\left\lbrace u_i^{(m)},\sigma_{i,j}^{(m)}\right\rbrace$ expressed in the coordinate system $(x_1,x_2,x_3)$ using the following equations:

\begin{equation}
\boldsymbol{u^{'(m)}}=R(\theta_m)\cdot\boldsymbol{u^{(m)}},\nonumber
\end{equation}
\begin{equation}
\boldsymbol{\sigma^{'(m)}}=R(\theta_m)\cdot\boldsymbol{\sigma^{(m)}}\cdot R(\theta_m)^T,\nonumber
\end{equation}

where $R(\theta)$ is the rotation matrix,
\begin{equation}
R(\theta)=\left[ \begin{array}{ccc} 
\cos(\theta) & 0 & -\sin(\theta)\\
0 & 1 & 0\\
\sin(\theta) & 0 &  \cos(\theta) \end{array}\right].\nonumber
\end{equation}

In order to define transmission and reflection coefficients, it is necessary to normalize each mode so that it carries unit energy flow through the interface: 
\begin{equation}
\begin{array}{lcr}
\boldsymbol{\overline{u}^{(n)}}=\dfrac{\boldsymbol{u^{(n)}}}{{C_n}} & \text{and} &\boldsymbol{\overline{\sigma}^{(n)}}=\dfrac{\boldsymbol{\sigma^{(n)}}}{{C_n}},\\ \nonumber
\end{array}
\end{equation}
$C_n$ being the normalization coefficient. To determine this coefficient, we use the bi-orthogonality relation established by Auld\cite{auld1973acoustic}, Fraser\cite{fraser1976orthogonality} and generalized by Gunawan\cite{gunawan2007reflection} that involves the bi-orthogonality coefficient $P_{mn}$ :
\begin{equation}
P_{mn}=\frac{i\omega}{4}\int_{-h}^{+h}\left[u_j^{\prime{(m)}}(\sigma_{1j}^{\prime{(n)}})^{*} - (u_j^{\prime{(p)}})^{*}\sigma_{1j}^{\prime{(m)}} \right]dx_2.\nonumber
\end{equation}
For propagating modes the coefficient $P_{mn}$ is non zero only when $m=n$. The real part of this coefficient corresponds to the energy flow passing through the interface. The coefficient $C_n$ for each propagating mode is thus given by :
\begin{equation}
C_n=Re\left\lbrace P_{n} \right\rbrace,
\label{NormaCoeff}
\end{equation}

For a non-propagating mode $m$, the normalization coefficient cannot be expressed using Eq. \eqref{NormaCoeff}, because the energy flow of this mode is by definition zero through the interface ($Re\left\lbrace P_{mm}\right\rbrace =0$). However, following Auld's work \cite{auld1973acoustic}, it exists for each non-propagating mode $m$ with a wavenumber $k_m$ a conjugate non-propagating mode $p$, associated with a wavenumber $k_p=k_m^*$. the combination of this modes gives rise to an energy flow given by the real part of $P_{mp}$. Each non-propagating mode $m$ can be normalized by this coefficient
\begin{equation}
C_m=Re\left\lbrace P_{mp} \right\rbrace,\nonumber
\end{equation}
In the following, $\overline{\boldsymbol{u}}$ and $\overline{\boldsymbol{\sigma}}$ will be written as $\boldsymbol{u}$ and $\boldsymbol{\sigma}$ to lighten the expressions.\\

The boundary conditions at the interface are the stress cancellation on the risers and the displacement and stress continuity on the central part. They can be written as:

\begin{equation}
{u}_j^{(I)}+\sum\limits_{n_1=1}^\infty r_{(I|n_1)}{\tilde{u}}_j^{(n_1)}=\sum\limits_{n_2=1}^\infty t_{(I|n_2)}{{u}_j^{(n_2)}}\text{ , } |x_2|<h_2,
\label{systcondcondiu}
\end{equation}

\begin{multline}
\sigma_{1j}^{(I)}+\sum\limits_{n_1=1}^\infty r_{(I|n_1)}{\tilde{\sigma}_{1j}^{(n_1)}}=\\ \left\lbrace \begin{array}{ll} 
0 & \text{ , } h_2<|x_2|<h_1\\
\sum\limits_{n_2=1}^\infty t_{(I|n_2)}{\sigma_{1j}^{(n_2)}} & \text{ , } |x_2|<h_2 \end{array}\right., \label{systcondcondis1i}
\end{multline}
with $j=1,2,3$. $r_{(I|n)}$ and $t_{(I|n)}$ represent the reflexion and transmission coefficients of the incident mode in the $n^{th}$ mode in the corresponding part of the plate. This system of equations cannot be solved analytically and it is necessary to truncate the series and discretize the displacement and stress fields.

\section{Inversion of the problem}

To solve numerically Eqs. \eqref{systcondcondiu}-\eqref{systcondcondis1i}, the stress and displacement fields need to be discretized along the normal to the plate with a thickness sampling pitch $\Delta x_2$. A maximum number of considered modes is then set by the following spatial Shannon criterion indicating that it is necessary to have at least two points by period:
\begin{equation}
k_{x_2} < \frac{2\pi}{\Delta{x_2}}\nonumber
\end{equation}
with $k_{x_2}=\sqrt{({\omega}/{c_T})^2-k^2}$. The number $N$ of selected modes is lower than the number of discrete points along the thickness. 
This discretization of the stress and displacement fields allows to write Eqs.\eqref{systcondcondiu}-\eqref{systcondcondis1i} in a matrix form :
 \begin{widetext}
 \begin{equation}
\underbrace{
\left(\begin{array}{c}
r_1 \\ \vdots \\ r_i\\ \vdots \\ r_{N_1} \\ \\ t_1\\ \vdots \\ t_i \\ \vdots \\ t_{N_2}
\end{array}\right)}_{\boldsymbol{C}}
=
{\underbrace{
\left(\begin{array}{ccccccccccc}
-\tilde{u}_1^{(1)} & & -\tilde{u}_1^{(n)}& & -\tilde{u}_1^{(N_1)}& & u_1^{(1)} & & u_1^{(n)} & &u_1^{(N_2)}\\
-\tilde{u}_2^{(1)}& &-\tilde{u}_2^{(n)} & & -\tilde{u}_2^{(N_1)}& &u_2^{(1)} & &u_2^{(n)}& & u_2^{(N_2)}\\
-\tilde{u}_3^{(1)} &  & -\tilde{u}_3^{(n)}& & -\tilde{u}_3^{(N_1)}& &u_3^{(1)}& &u_3^{(n)} &  & u_3^{(N_2)}\\
 &\cdots & &\cdots & & & &\cdots & &\cdots & \\
-\tilde{\sigma}_{11}^{(1)}& &-\tilde{\sigma}_{11}^{(n)} & & -\tilde{\sigma}_{11}^{(N_1)} & & \sigma_{11}^{(1)} & & \sigma_{11}^{(n)}& &\sigma_{11}^{(N_2)}\\
-\tilde{\sigma}_{12}^{(1)}& &-\tilde{\sigma}_{12}^{(n)}& & -\tilde{\sigma}_{12}^{(N_1)} & & \sigma_{12}^{(1)} & & \sigma_{12}^{(n)}& &\sigma_{12}^{(2,N_2)}\\
-\tilde{\sigma}_{13}^{(1)}& & -\tilde{\sigma}_{13}^{(n)}& & -\tilde{\sigma}_{13}^{(N_1)} & & \sigma_{13}^{(1)} & & \sigma_{13}^{(n)}& & \sigma_{13}^{(N_2)}
\end{array}\right) \\}_{\boldsymbol{M}}}^{-1}
\cdot
\underbrace{
\left(\begin{array}{c}
u_1^{(I)}\\u_2^{(I)}\\u_3^{(I)}\\ \\ \sigma_{11}^{(I)}\\ \sigma_{12}^{(I)} \\ \sigma_{13}^{(I)}
\end{array}\right)}_{\boldsymbol{Y}}.
 \end{equation}
\end{widetext}
$\boldsymbol{C}$ is the vector of all the reflection and transmission coefficients, $\boldsymbol{M}$ is the matrix containing the displacement-stress field of each mode, and $\boldsymbol{Y}$ is the displacement-stress field of the incident mode.
The rectangular matrix $\boldsymbol{M}$ is inverted using a Moore-Penrose pseudo-inversion. The chosen sampling interval $\Delta{x_2}=2\cdot 10^{-4}$ mm implies the consideration of $N_L=241$ Lamb modes and $N_{SH}=120$ SH modes in each part of the plate. This choice is made in order to fulfill the energy conservation condition with a reasonable precision such that :
$1 - (\sum\limits_{i=1}^N |r_i|^2+|t_i|^2) < 10^{-3}$.

\section{Optimization of the negative refraction phenomenon}

\begin{figure}
  \centering
\includegraphics[width=\columnwidth]{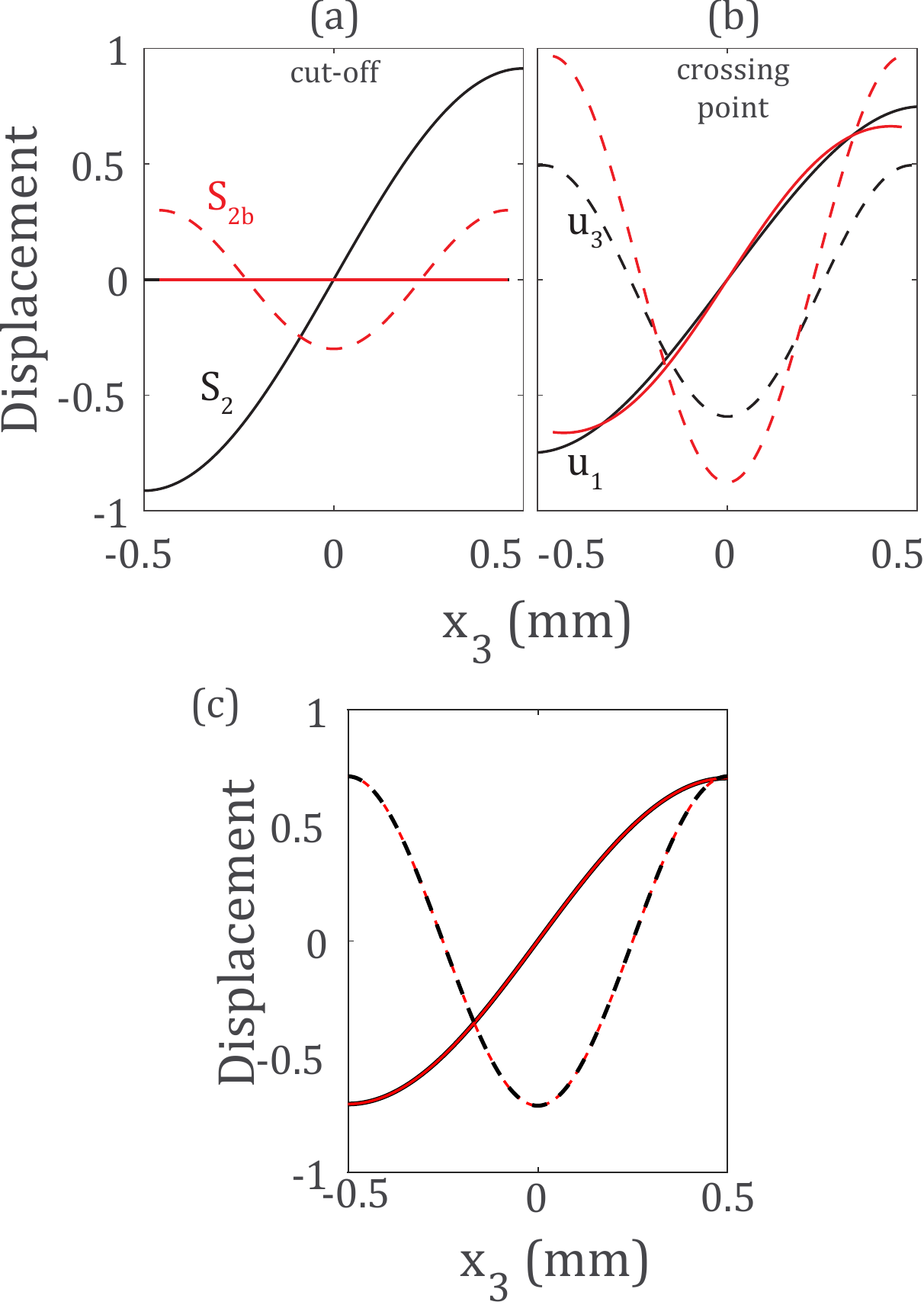}
\caption{\label{fig:displacementCrossCut}Displacement $u_3$ (continuous line) and $u_1$ (dotted line) of $S_2$ (black) and $S_{2b}$ (red) in a $1$ mm thick Duralumin plate : (a)  at the cut-off frequency. (b) At the crossing point. (c) At the coincidence when $\nu=1/3$}
\end{figure}

This semi-analytical model is first used to determine the thickness ratio that maximizes the conversion between the forward mode $S_2$ and the backward one $S_{2b}$ at normal incidence.
\begin{figure}[tb]
\centering
\includegraphics[width=\columnwidth]{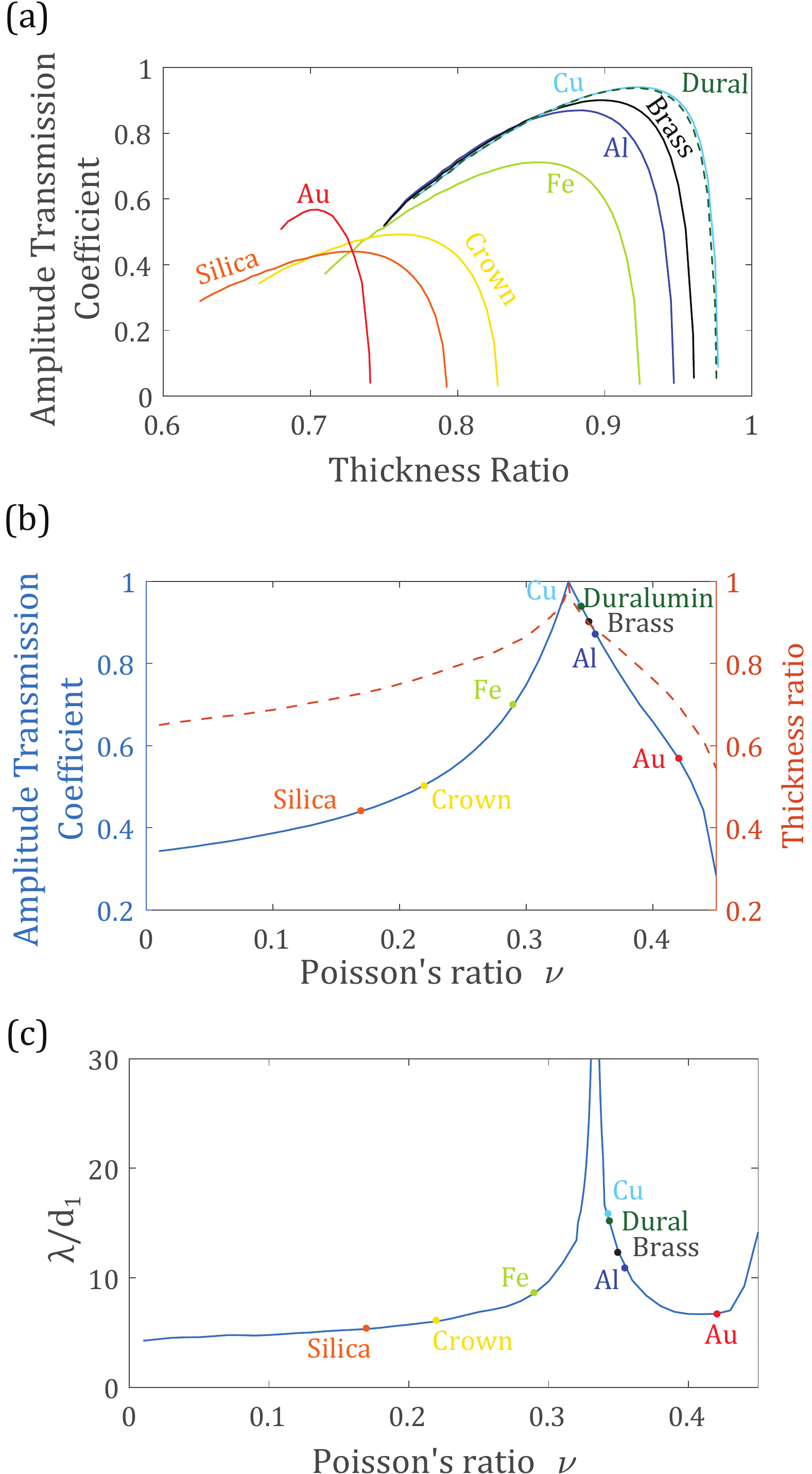}
\caption{\label{fig:opth2mat_nu} Transmission of the $S_2$ mode at normal incidence: (a) Transmission coefficient $|t_{S_2 \to S_{2b}}|$ as a function of the thickness ratio $d_2/d_1$ for different materials. (b) Amplitude transmission coefficient (continuous line) and the associated best thickness ratio (dotted line) as a function of the Poisson's ratio. The coefficient reaches $1$ for $\nu=1/3$, that is to say when $S_2$ and $S_{2b}$ share the same cut off frequency. (c) Evolution of the crossing wavelength as a function of the Poisson's ratio.}
\end{figure}
Figure \ref{fig:opth2mat_nu}(a) displays the transmission coefficient $|t_{S_2\to S2b}|$ at normal incidence for various materials as a function of the thickness ratio $d_2/d_1$. For each thickness ratio, the amplitude transmission coefficient is calculated at the crossing frequency, intersection of $S_2$-mode in the thick part and $S_{2b}$-mode in the thin part [see Fig. \ref{fig:stepschema_curve}(b)]. Interestingly, the amplitude transmission coefficient can be close to unity for materials such as Duralumin or Copper. This can be explained by the close displacement profiles of the two modes at the crossing frequency [see Fig. \ref{fig:displacementCrossCut}(b)]. However, for each material, the amplitude transmission coefficient strongly decreases when the thickness ratio tends towards unity. In that asymptotic case, the crossing frequency approaches the cut-off frequencies where the $S_2$-mode tends to be purely longitudinal while the $S_{2b}$-mode becomes purely shear. An important mode mismatch is thus found when $d_2/d_1 \rightarrow 1$ [see Fig. \eqref{fig:displacementCrossCut}.(a)]. Figure \ref{fig:opth2mat_nu}(b) displays $|t_{S_2\to S2b}|$ and the optimum thickness ratio as a function of the Poisson's ratio $\nu$. Interestingly, when $\nu$ tends to the value $1/3$, the amplitude transmission coefficient reaches unity with an optimum thickness ratio of 1. This critical value of $\nu$ indeed implies the coincidence of $S_2$ and $S_{2b}$ cut-off frequencies. $S_2$ and $S_{2b}$ modes are thus strictly identical in that case, which means a full mode overlap and a perfect conversion between them (see Fig. \ref{fig:displacementCrossCut}.c). Such case has been recently investigated by Stobbe \textit{et al}\cite{stobbe2017conical} as it also gives rise to a Dirac cone in the dispersion curves. This means that the group velocity remains finite while the wave number tends to zero. However, in the present case, an infinite wavelength limits the experimental interest for this ideal case. A compromise has thus to be found between the transmission coefficient and the mode wavelength and in that respect, the choice of Duralumin appears to be optimal: the transmission coefficient reaches $|t_{S_2\to S2b}|=0.94$ for a thickness ratio $d_2/d_1$ of $0.92$ and a reasonable wavelength $\lambda =15 d_1$. Moreover, Duralumin has a much lower absorption coefficient than Copper ($\sim 1$ dB/m for Duralumin and in that respect, $\sim 20$ dB/m for Copper \cite{prada2008power}).

\begin{figure}
\centering
\includegraphics[width=\columnwidth]{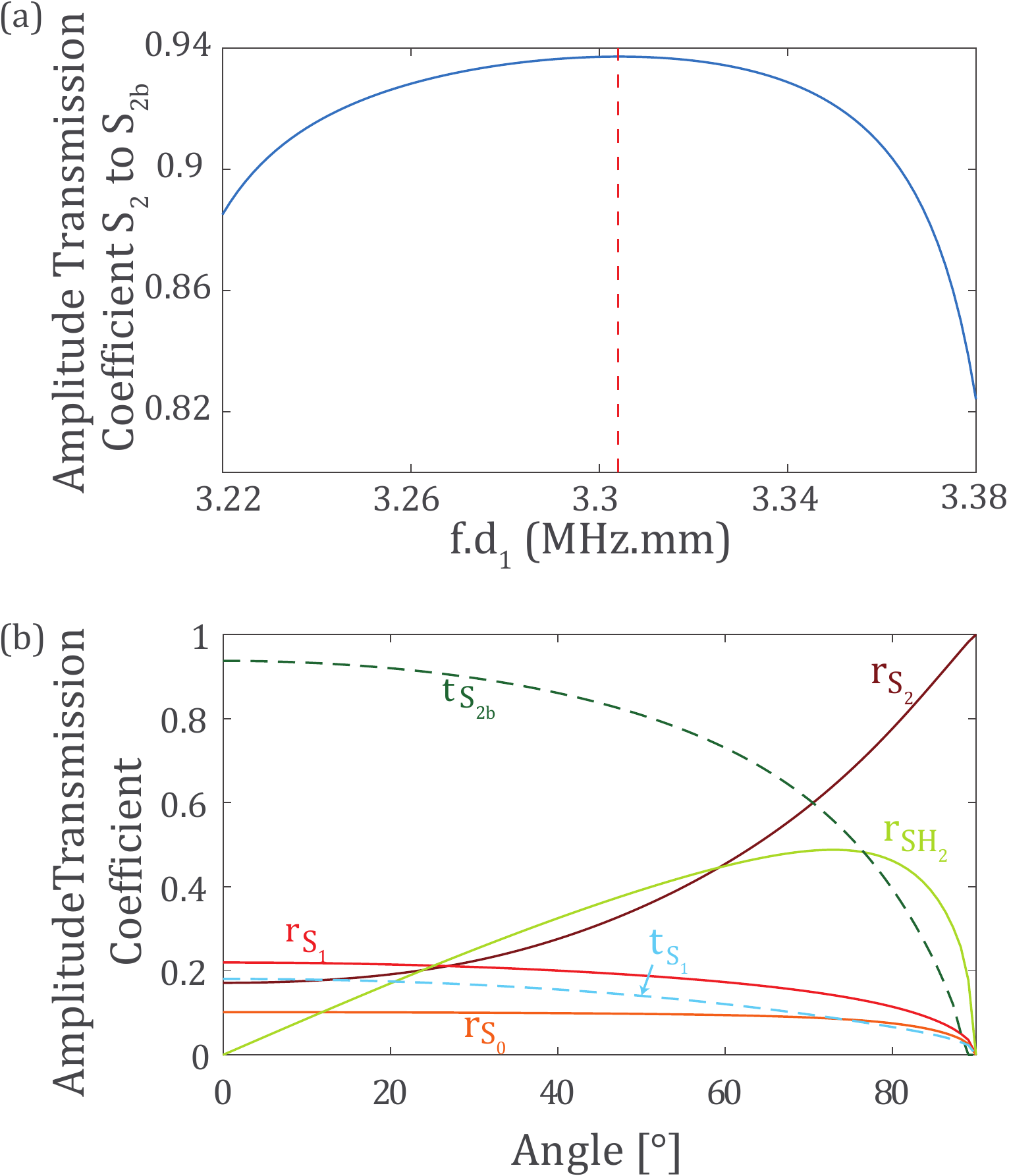}
  \caption{\label{fig:optf0_angl}(a) Variation of the amplitude transmission coefficient at the optimum thickness ratio as a function of frequency. (b) Reflection and transmission coefficients as a function of the $S_2$-mode incident angle at frequency $f=3.33$ MHz. (The $S_2$ to $SH_0$ reflection coefficient and the transmission ones to $S_0$ and $SH_0$ are not shown here because smaller than $10^{-3}$)}
\end{figure}

Now that the Duralumin has been chosen, we investigate the bandwidth over which the conversion between $S_2$ and $S_{2b}$ remains efficient.
The frequency dependence of $|t_{S_2\to S2b}|$ is displayed in Fig. \ref{fig:optf0_angl}(a). The negative refraction of Lamb waves appears to be broadband: for $d_1=1$mm, the transmission coefficient is above $0.9$ over a frequency bandwidth $\Delta f \sim 0.15$ MHz. The negative refraction phenomenon can thus be observed in the time domain for wave-packets of length $\Delta t \sim 1/\Delta f \sim 6$ $\mu$s. This important feature will be confirmed experimentally in the next section.

The angular dependence of the negative refraction phenomenon is also particularly important for the implementation of a negative refraction flat lens. Figure \ref{fig:optf0_angl}(b) displays the reflection and transmission coefficients for the various propagating  modes supported by each part of the plate for an incident $S_2$ mode. It appears that $|t_{S2->S2b}|$ remains above $0.8$ over an angular range of $45^{\circ}$. 
Note that, for large angles of incidence, the $S_2$-mode is mainly reflected into itself and the $SH_2$-mode. As we will see now, this angular robustness of the $S_2\rightarrow S_{2b}$ conversion ensures a large aperture angle for the negative refraction flat lens. 

\section{Negative refraction lens}

\begin{figure}[t!]
  \centering
  \includegraphics[width=\columnwidth]{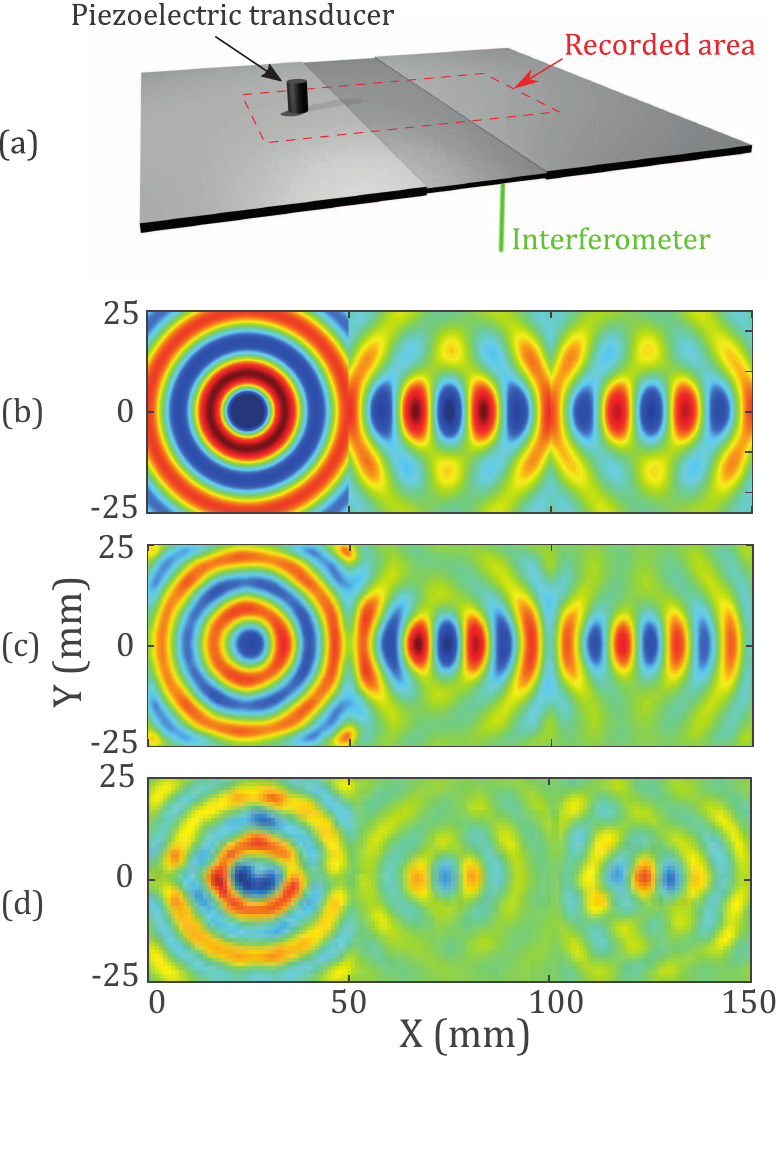}
  \caption{\label{fig:compalentille} (a) Experimental Setup. Wave-field associated with the $S_2$ and $S_{2b}$ modes supported by the NR lens made of a Duralumin plate with a thickness ratio $d_1/d_2$ of $0.9$.  (b) Semi-analytical result. (c) FDTD numerical result. (d) Experimental result obtained with laser interferometry.}
\end{figure}

Now that the angular and temporal stability of the negative refraction phenomenon has been investigated, the negative refraction flat lens is investigated theoretically, numerically and experimentally. A negative refraction lens for Lamb waves consists in a plate with the association of a downward step and an upward step. In the last section, only the conversion at a downward step has been considered. According to the reciprocity principle, the transmission coefficient from $S_{2b}$ to $S_2$ at an upward step is strictly identical to the  transmission coefficient from $S_2$ to $S_{2b}$ at a downward step. Hence, the wave-field generated by a point source at the input of a negative refraction lens [Fig. \ref{fig:compalentille}(b)] can be easily predicted. The angular stability of the negative refraction phenomenon at a thickness step previously observed in Fig. \ref{fig:optf0_angl}(b) give rise to excellent focusing properties for the negative refraction flat lens, at least theoretically. These theoretical predictions are confirmed in Fig. \ref{fig:compalentille}(c) that shows the result of a numerical simulation performed with a finite-difference time domain software on the same device\cite{bossy2004three}. A close agreement is found between the semi-analytical result and the numerical simulation, which confirms the validity of our approach. \\ 
The optimized negative refraction lens is now implemented experimentally. It consists of a $1$-mm-thick Duralumin plate that has been engraved by chemical erosion using iron chloride to obtain a $0.9$ mm thick thin part. The plate dimensions are chosen such that the reflections on the free edges of the plate are limited during the recording. The excitation of the plate is achieved by a piezoelectric transducer (Olympus V183-RM) of $10$ mm-diameter glued (with salol or phenyl salicylate) on the thick part at a distance $D=25$ mm from the step. A 5 $\mu s$ chirp signal spanning the frequency range $3.05$-$3.65$ MHz is sent to the transducer which generates a cylindrical incident wavefront in the plate. The out-of-plane displacement is measured with a photorefractive interferometer (from BossaNova, Tempo1D) over a grid of points that maps $150$ $\times$ $50$ mm$^2$ of the plate surface, with $1$ mm-pitch across the thin part. Signals detected by the optical probe are fed into a high speed usb oscilloscope (TiePie HS5) and transferred to a computer. A spatio-temporal discrete Fourier transform (DFT) of the recorded wavefronts is then performed from $3.22$ to $3.52$ MHz and for spatial frequencies $k/(2\pi)$ ranging from $-0.15$ to $0.15$ mm$^{-1}$. Figure \ref{fig:compalentille}(d) shows the normal displacement field measured on the plate filtered at the frequency $f=3.33$ MHz. In order to avoid the reflections on the free edges of the plates, the DFT is calculated for adapted time windows in each part of the lens : $0-20$ $\mu$s for the first thick part, $10-70$ $\mu$s for the thin part and $40-100$ $\mu$s for the second thick part. Despite the inherent imperfections to the experimental realization, the measured wave field is quite remarkably congruent with the theoretical prediction and the numerical simulation. The angular spectrum of the negative refraction observed on Fig. \ref{fig:compalentille}.d is really alike the one theoretically predicted ($45^\circ$, cf Fig. \ref{fig:optf0_angl}.b). 
The ultrasound laser experiments also allows to investigate the behavior of the plate lens in the time domain\cite{philippe2015focusing}. Due to the spectral robustness of the $S_2\rightarrow S_{2b}$ conversion [Fig. \ref{fig:optf0_angl}.(a)], the plate lens also operates in the time domain for wave packets of finite duration ($6$ $\mu$s). The result is displayed in the Supplementary movie\cite{SeeSuppl}.\\

\section{Conclusion}

A semi-analytical study of the interaction of Lamb waves with a thickness step-like and more specifically the conversion between forward and backward Lamb modes, associated with negative refraction was proposed. The semi-analytical model allows to investigate this phenomenon by computing the transmission and reflection coefficient as a function of Poisson's ratio, thickness ratio, angle of incidence and frequency. Then find the optimal design to achieve negative refraction. This semi-analytical model predicts the frequency and angular robustness of the negative refraction process and shows that a Duraluminium plate as used in previous works \cite{philippe2015focusing,prada2005laser} was an excellent compromise to observe negative refraction. Relying on these results, we have designed and tested a negative refraction lens acting in the time domain. The wave-field recorded by means of laser ultrasound techniques is in excellent agreement with theoretical predictions and FDTD numerical simulations. The perspective of this work will be to investigate negative refraction related phenomena such as the notion of complementary media\cite{pendry2003focusing} and the ability to cancel the propagation of waves by adjoining two mirror regions of opposite refractive indices. Beyond negative refraction, the proposed theoretical model is much more general and can be applied to all kind of discontinuity and Lamb modes. The consideration of evanescent or inhomogeneous Lamb modes in the model also paves the way towards the implementation of a superlens for Lamb waves.\\

\section*{Acknowledgments}

The authors are grateful for funding provided by the Agence Nationale de la Recherche (ANR-15-CE24-0014-01, Research Project COPPOLA)and by LABEX WIFI (Laboratory of Excellence within the French Program Investments for the Future, ANR-10-LABX-24 and ANR-10-IDEX-0001-02 PSL*). B.G. acknowledges financial support from the French ``Direction Générale de l'Armement'' (DGA).



\end{document}